\documentclass[11pt,twoside]{article}

\usepackage{asp2006}
\usepackage{epsf}
\usepackage{lscape}
\usepackage{graphicx}

\begin{document}
\title{Synchronization of the G Giant Rotation in the Symbiotic Binary StH$\alpha$ 190?}
%\title{Possible pulsations of the G giant in the symbiotic binary StH$\alpha$ 190}   %%% Fill in title
\author{M. Cika\l{}a$^1$, M. Miko\l{}ajewski$^1$, T. Tomov$^1$, D. Kolev$^2$, L. Georgiev$^3$, U. Munari$^4$, P. Marrese$^5$, T. Zwitter$^6$}
\affil{$^1$Centre for Astronomy of Nicolaus Copernicus University, Toru\'n, Poland}    %%% Fill in author affiliations
\affil{$^2$National Astronomical Observatory Rozhen, Institute of Astronomy, Smolyan, Bulgaria}
\affil{$^3$Instituto de Astronom\'{i}a, UNAM, M\'{e}xico DF, M\'{e}xico}
\affil{$^4$INAF, Osservatorio Astronomico di Padova, Sede di Asiago, Italy}
\affil{$^5$Leiden Observatory, Leiden, The Netherlands}
\affil{$^6$University of Ljubljana, Faculty of Mathematics and Physics, Ljubljana, Slovenia}

\begin{abstract}
We present an analysis of high resolution spectral observations of the symbiotic star StH$\alpha$ 190. A 30 days period has been derived from radial velocities of the {\it G-type} absorption lines and the HeII~$\lambda4686$\AA\, emission line. The main aim of this work was to look for explanation of the very wide absorption lines of the yellow giant. The very low mass function obtained from the absorption lines radial velocities suggests that the observed changes probably do not corespond to the orbital motion of this star.
\end{abstract}

\section{Background, observations and methods}
\begin{figure}[!t]
\begin{center}
\includegraphics[width=0.5\textwidth]{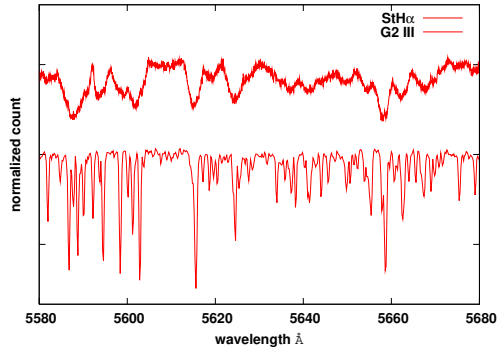}
\caption{Apart of the optical spectrum of StH$\alpha$ 190 the spectrum of the standard star \textit{G2 III (HD126868)} from \textit{OHP} \textsc{ellodie} library is plotted for comparision.
}
\label{spectrum}
\end{center}
\end{figure}

StH$\alpha$ 190 was discovered by \citet{stephenson}, during an objective-prism survey for emission-lines objects. Now it is classified as a \textit{yellow} $d'$ symbiotic star. The system consist with a rapidly rotating (Figure \ref{spectrum}) $v_{rot}=105 km\,s^{-1}$ \citep{munari01} \textit{G2 III/IV} star \citep{smith}, and a moderate temperature $T_h\sim50000$ white dwarf \citep{schmid}.

An inspection of the catalogue of rotational velocities (de Medeiros et al. 1999) shows that almost all G stars rotate with $v_{rot}<10\,km\,s^{-1}$, and only eleven of listed stars have $v_{rot}>20\,km\,s^{-1}$. Most of these relatively fast rotating stars are recognized as peculiar or ''single-lined'' spectroscopic binaries.

Periodic changes of about 900 days and an amplitude 0.16 mag, were discovered in \textit{K-band} photometry by \citet{whitelock}. They speculated that tis is system orbital period. \citet{munari01} suggest a possible 171 days spectroscopic period of StH$\alpha$ 190, based on \textit{NaD} radial velocities. \citet{smith} showed that a shorter 37-39 days period is more possible.

We have used 30 high resolution spectra with resolving power $16000<R<54000$ obtained during four years. The spectra have been processed in a standard \textsc{iraf} way. To measure the radial velocities of emission lines, we used the \textsc{iraf}'s RV package. For the absorption lines radial velocity measurements the cross-correlation method has been used. As a reference standard spectrum we chose a \textsc{G2 III} one from the synthetic spectra library \citet{zwitter}, on basis on the parameters obtained by \citet{smith}.

The Fast Fourier Transform (FFT) for unequally spaced data was computed for a periodicity analysis. The simply \citet{deeming} method has been used. Using the mean square method, the simple sinus function has been fitted to phased data.

%\begin{figure}[!t]
%\begin{center}
%\includegraphics[width=0.9\textwidth]{cikala2fig3.eps}
%\epsfig{file=K.eps,height=0.6\textheight,angle=-90}
%\caption{The \textit{K-band} photometry. Left panel shows the FFT of K magnitudes, right one phased data with period %$P=905.^{d}57$, which corresponds to the biggest peak at power window of FFT graph.}
%\end{center}
%\end{figure}

\section{The spectroscopic period}

\begin{figure}[!t]
\begin{center}
\includegraphics[width=0.85\textwidth]{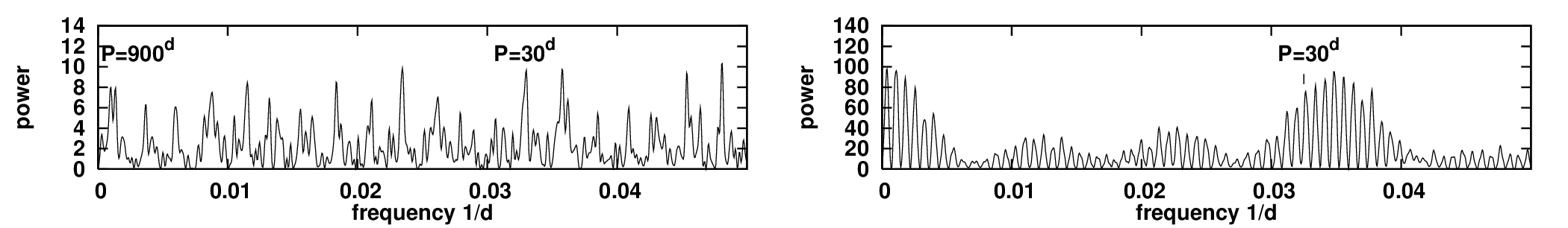}
\includegraphics[width=0.85\textwidth]{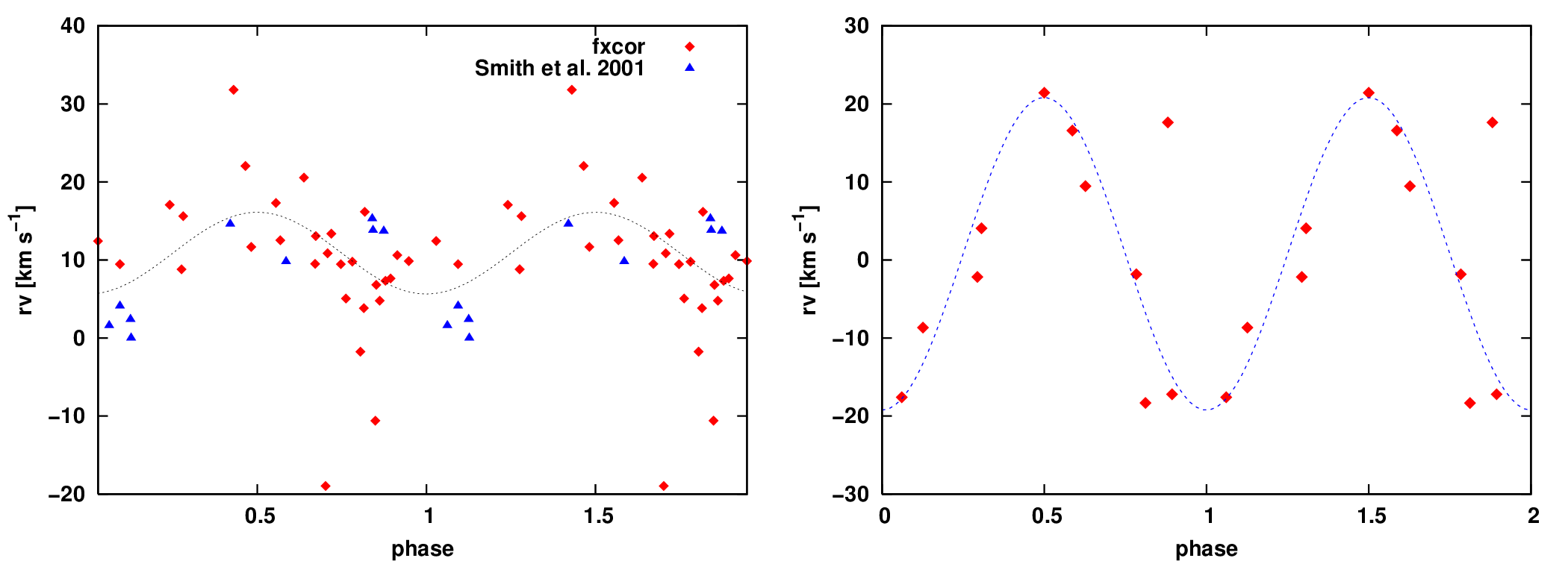}
\caption{Upper row presents the FFT of radial velocites of \textit{G2} star absorption lines (left) and HeII~$\lambda4686$\AA\,  emmision line (right). Lower row shows phased radial velocities of mentioned lines with period $P=30.^{d}488$ and $P=30.^{d}497$ respectively for cool star absorption lines and HeII~$\lambda4686$\AA\, emission lines. The moments of conjunction ($\phi=0.25$) are almost the same. The radial velocities of mass centres differ of about $10\,km\,s^{-1}$. 
}
\label{spec}
\end{center}
\end{figure}

Figure~\ref{spec} shows the radial velocities \textit{FFT} of the {\it G-type} star absorption lines and the HeII~$\lambda 4686$\AA\, emission line, as well as the phased data. The detailed analysis of the periodograms showed that the main period of changes observed in radial velocity is 30.5 days in both cases. The HeII~$\lambda4686$\AA\, emission line was presented only in eleven spectra during the four years observing period. The spectral windows of the Fourier transforms obtained from the radial velicities of the \textit{G2} star absorption lines, and the HeII~$\lambda 4686$\AA\, emission line are very different. The fit details are listed in Table~\ref{rvsol}.

\begin{table}[!b]
\begin{center}
\caption{Summary of radial velocity changes. 'P' is the orbital period, 'K' is the semi-amplitude, $\gamma$ - radial velocity of mass center, and $T_{0}$ - moment of conjuction.}
\begin{tabular}{|c|c|c|c|c|}
\hline
 & $P\, [d]$ & $K\,[km\,s^{-1}]$ & $\gamma\,[km\,s^{-1}]$ & $T_{0}\,[HJD]$ \\%
\hline
\hline
Absorptions & $30.488$ & $5.238$ & $10.876$ & $2451564.922$\\
HeII~$\lambda4686$\AA & $30.497$ & $20.035$ & $0.790$ & $2451564.189$\\
\hline
\end{tabular}
\label{rvsol}
\end{center}
\end{table}

\section{Spectral Energy Distribution}

\begin{figure}[!t]
\begin{center}
\includegraphics[width=0.65\textwidth]{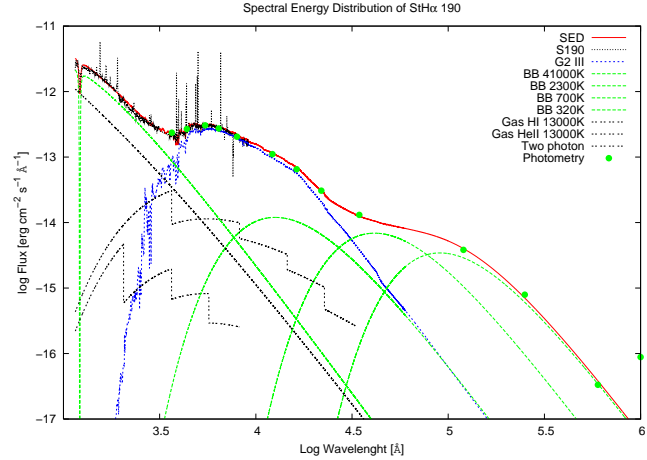}
\caption{Spectral energy distribution of StH$\alpha$ 190 from UV to IRAS 120$\mu m$ band.}
% Hot stellar source is represented by \textit{black body}. The Raileigh scattering process have been included. The recombination continuum is express by both HI end HeII, and a two-photon emission continuum. The model of \textsc{G2 III} star was taken from 1\AA\, spectral liblary (Munari et al. 2005). The infrared radiation consist three \textit{black body} model.}
\label{sed}
\end{center}
\end{figure}

Following \citet{skopal} we have calculated the spectral energy distribution (see Figure~\ref{sed}). The model includes: the hot and the cool stars, the gaseous nebula, and the dust radiations. We have completed the model of the hot gas radiation with a two photon radiation ($2\gamma$), which could be a significant source of the observed ultraviolet flux. The model of \textsc{G2 III} star was taken from 1\AA/pix spectral library of Munari et al. (2005). The infrared radiation is fitted with three black bodies.

\section{Summary}

The spectral energy distribution (Figure \ref{sed}) confirms the presence of a \textit{G2 III} star in the system. We have obtained $R\sim10R_{\odot}$ and $L\sim63L _{\odot}$, in agrement with the values $R\sim8R_{\odot}$ and $L\sim50L _{\odot}$
reported by Smith et al. (2001).

Using the period of 30.5 days and the semi-amplitude of the radial velocity changes of about $5.2\,km\,s^{-1}$ we obtained a mass function $f(m)=0.0005\,M_{\odot}$. Such low mass function indicates very high mass ratio of about $q\geq15$. Using for the cool star a mass $2.5M_{\odot}$ we can estimate lower value for mass of the invisible companion as $M_h \sin{i} \leq 0.15 M_{\odot}$. Such mass of the hot companion is unrealistic which suggest that radial velocities variations do not reflect the orbital motion. This means that 30 days period is not orbital one and that we cannot consider the synchronous rotation of the yellow giant as an explanation of the very wide absorption lines in its spectrum.

A more detailed analyses of the absorption lines in the observed in optical spectrum of StH$\alpha$ 190 are needed to solve the problem with observed high rotation of the yellow giant.

\acknowledgements
This work was supported by the Polish MNiSW Grant N203 018
32/2338. Partly based on observations made with the European Southern Observatory telescopes
obtained from the ESO/STECF Science Archive Facility and spectral data retrieved from the ELODIE
archive at Observatoire de Haute-Provence (OHP).

\end{document}